\documentclass[conference]{IEEEtran}
\usepackage{cite}
\usepackage{url}

\ifCLASSINFOpdf
 \usepackage[pdftex]{graphicx}
\else

\fi

\usepackage{subfigure}

\usepackage{algorithmic}
\usepackage{algorithm}  
\usepackage{color}
\usepackage{xcolor}

\makeatletter
\newcommand\fs@betterruled{%
	\def\@fs@cfont{\bfseries}\let\@fs@capt\floatc@ruled
	\def\@fs@pre{\vspace*{5pt}\hrule height.8pt depth0pt \kern2pt}%
	\def\@fs@post{\kern2pt\hrule\relax}%
	\def\@fs@mid{\kern2pt\hrule\kern2pt}%
	\let\@fs@iftopcapt\iftrue}
\floatstyle{betterruled}
\restylefloat{algorithm}
\makeatother



\usepackage{amsmath,amsthm,bm,amssymb}
\newtheorem{defi}{Definition}
\newtheorem{prop}{Proposition}

\newtheorem{thm}{Theorem}
\newtheorem{ass}{Assumption}

\IEEEoverridecommandlockouts
\begin{document}

\title{Insurance Contract for High Renewable Energy Integration}

\author{Dongwei Zhao, Hao Wang,  Jianwei Huang, Xiaojun Lin

\thanks{This work is supported in part by the Shenzhen Science and Technology Program (Project JCYJ20210324120011032), Guangdong Basic and Applied Basic Research Foundation (Project 2021B1515120008),  the Shenzhen Institute of Artificial Intelligence and Robotics for Society,  NSF through grant ECCS-2129631, and the Australian Research Council (ARC) Discovery Early Career Researcher Award (DECRA) under Grant DE230100046.
}
\thanks{Dongwei Zhao is with the MIT Energy Initiative, Massachusetts Institute of Technology. Hao Wang is with the Department of Data Science and Artificial Intelligence, Monash University. Jianwei Huang is with the School of Science and Engineering, The Chinese University of Hong Kong, Shenzhen and  Shenzhen Institute of Artificial Intelligence and Robotics for Society (corresponding author). 
Xiaojun Lin is with the Elmore Family School of Electrical and Computer Engineering, Purdue University.}
}

\IEEEoverridecommandlockouts

\IEEEpubid{\makebox[\columnwidth]{978-1-6654-3254-2/22/\$31.00~\copyright2022 IEEE \hfill}\hspace{\columnsep}\makebox[\columnwidth]{ }}

\maketitle
\IEEEpubidadjcol

\begin{abstract} 
The increasing penetration of renewable energy poses significant challenges to power grid reliability. There have been increasing interests in utilizing financial tools, such as insurance, to help end-users hedge the potential risk of lost load due to renewable energy variability. With insurance, a user pays a premium fee to the utility, so that he will get compensated in case his demand is not fully satisfied. 
A proper insurance design needs to resolve the following two challenges: (i) users' reliability preference is private information; and (ii) the insurance design is tightly coupled with the renewable energy investment decision. To address these challenges, we adopt the contract theory to elicit users' private reliability preferences, and we study how the utility can jointly optimize the insurance contract and the planning of renewable energy. A key analytical challenge is that the joint optimization of the insurance design and the planning of renewables is non-convex. We resolve this difficulty by revealing important structural properties of the optimal solution, using the help of two benchmark problems: the no-insurance benchmark and the social-optimum benchmark. Compared with the no-insurance benchmark, we prove that the  social cost and  users' total energy cost are always no larger under the optimal contract.  Simulation results show that the largest benefit of the insurance contract is achieved at a medium electricity-bill price together  with a low type heterogeneity and a high renewable uncertainty.

\end{abstract}

\IEEEpeerreviewmaketitle

\section{Introduction}
\subsection{Background and motivation}

The transition from fossil energy sources to renewable energy sources is accelerating at the global scale  as the threat of climate change continues to grow. From 2014 to 2020, the globally installed capacity of solar PV has increased from about 40 GW to 140 GW, while the wind energy capacity has seen a growth from about 50 GW to 90 GWh \cite{renreport2021}. Many countries have set sustainability goals to increase the penetration of renewable energy in electricity supplies. For example, the California electricity market in the US sets a target of generating over two-thirds of its electricity from non-fossil and non-nuclear resources by 2030 \cite{targetrenca}. 

The increasing penetration of renewable energy generation, however, poses challenges to system reliability. When a significant amount of renewable energy supply fluctuates based on weather conditions, users may experience lost loads (i.e., unsatisfied demands) more often \cite{denholm2021challenges}. In principle, lost load could be alleviated by investing in other costly low-carbon energy resources. For example, energy storage could provide flexibility, which is however costly. Alternatively, the utility can simply invest in more renewable capacity to reduce the probability of low supply, which can also be costly. Therefore, it is crucial to provide the right financial arrangement so that not only can end-users hedge the potential risk due to lost load, but the utility is also incentivized to invest in the right amount of additional resources to improve grid reliability. 

The traditional financial arrangement between the utility and the end-users is usually through the electricity bill at either static or dynamic rates. Unfortunately, such an electricity bill does not provide a way for the end-users to get compensation for lost load, nor does it enable the utility to directly recover the cost of reliability-improving investment. In contrast, in this paper we are interested in more advanced financial tools, in particular insurance, to address the above open question. With insurance, a user pays a premium fee to the utility in advance, so that he will get compensated if his demand is not fully satisfied \cite{fuentes2019vertical}. In return, the utility can use the premium fees to invest in reliability-improving resources. Note that by choosing among different insurance contracts, the end-users can express their heterogeneous valuation of lost load. Therefore, we expect that the insurance contract will align the utility's cost with end-users' value of lost load more accurately, so that the utility is incentivized to invest in the right amount of additional resources. As a first step towards this direction, in this paper, we focus on a very-high renewable penetration setting, where the sole investment option that the utility considers is to invest in more renewable generation.  We leave as future work the study of the investment of other resources, such as energy storage.

To design a proper insurance scheme, however, the utility needs to resolve two challenges.  First, different users have distinct preferences of reliability \cite{ovaere2019detailed}, which will influence the type of insurance that they want to purchase.  However, such user preferences are often private information unavailable to the utility. Second, since the utility's renewable energy investment can affect the system reliability, it will also impact users' insurance choices.  Hence, it is crucial for the utility to jointly consider the renewable energy planning and insurance design together. These challenges motivate us to answer the following key question in the scenario of very-high renewable energy penetration: 

\begin{itemize}
\item How to jointly optimize an insurance contract and the renewable energy planning, which helps users hedge the risks of lost load and helps the utility  improve social welfare?
\end{itemize}

\subsection{Related works}

Several recent work has started to explore the benefits and design of reliability insurance for electricity users. Billimoria \textit{et al}. in \cite{billimoria2019market} and Fuentes \textit{et al}. in \cite{fuentes2019vertical} introduced the concepts of reliability insurance for end users. However, these studies did not provide quantitative results in the insurance design. 

Regarding insurance design using quantitative methods, Fuentes \textit{et al}. in \cite{fuentes2020using} conducted numerical studies to demonstrate the benefits of insurance for users, assuming that users have homogeneous preferences towards reliability.  Abedi \textit{et al}. in \cite{abedi2013cdf}, Dong \textit{et al}. in \cite{Dong2019insurance}, Niromandfam \textit{et al}. in \cite{niromandfam2020design}, and Fumagalli \textit{et al}. in \cite{fumagalli2004quality} considered users' heterogeneity in reliability preference and risk aversion in the insurance design. However, these studies did not consider the planning of generation resources. As we mentioned earlier, since the electricity supply can affect the system reliability and users' energy costs, it is important to consider the coupling  between insurance design and renewable energy investment. 

The most relevant work to our study is   \cite{billimoria2021design}, where Billimoria \textit{et al}. studied users' heterogeneity in reliability preferences and optimized investment decisions of electricity suppliers. However, the authors in \cite{billimoria2021design} assumed that users would voluntarily express to the insurance provider their reliability preferences and  compensation values, which is a rather unrealistic assumption given users' private information. 

None of the above studies considered the joint optimization of insurance design and renewable energy planning under users' private reliability preference. Thus, our work represents an important step towards filling this gap, by adopting contract theory to study such a joint optimization.

\subsection{Main results}

Focusing on an extreme future scenario of 100\% renewable energy generation, we summarize the key results and contribution of this paper as follows: 

\begin{itemize}

\item  \textit{Insurance contract for reliability:} To the best of our knowledge, our work is the first to jointly design the insurance contract for end-users' lost load  and optimize the planning of renewable energy. Such an insurance scheme can play an important role in the  future scenario of high renewable energy integration.

\item  \textit{Contract design and customized solution:} We adopt contract theory to solve the challenge of eliciting users' private preferences of reliability. Due to the use of contract theory, however, the joint optimization of contract design and renewable energy planning involves solving a challenging non-convex optimization problem. To resolve this difficulty, we characterize important structural properties of the optimal solution, using the help from two benchmark problems: the no-insurance benchmark and the social-optimum benchmark.

\item  \textit{Benefits of insurance contract:} 
Compared with the no-insurance benchmark,  we prove that the optimal insurance contract will never reduce social welfare or increase the users' total energy cost. Simulation results show that the optimal insurance contract reduces the social cost and users' costs by up to 20\% and 19\%, respectively. 

\item  \textit{Impact of electricity-bill price, type heterogeneity, and renewable uncertainty}:  The specific level of insurance benefits in reducing user cost and improving social welfare is affected by a number of factors.  Simulation results show that the highest benefit  is achieved at a medium electricity-bill price together with a low heterogeneity of types and a high uncertainty of renewables. 
\end{itemize}

\section{System Model}
  
We consider one electricity utility serving a group of $\mathcal{I}=\{1,2\ldots I\}$ users. Besides charging users electricity bills, the utility provides a set of insurance contract items for users. Users can purchase insurance to hedge the risk of lost load.

\begin{figure}[t]
	\centering
	\includegraphics[width=3in]{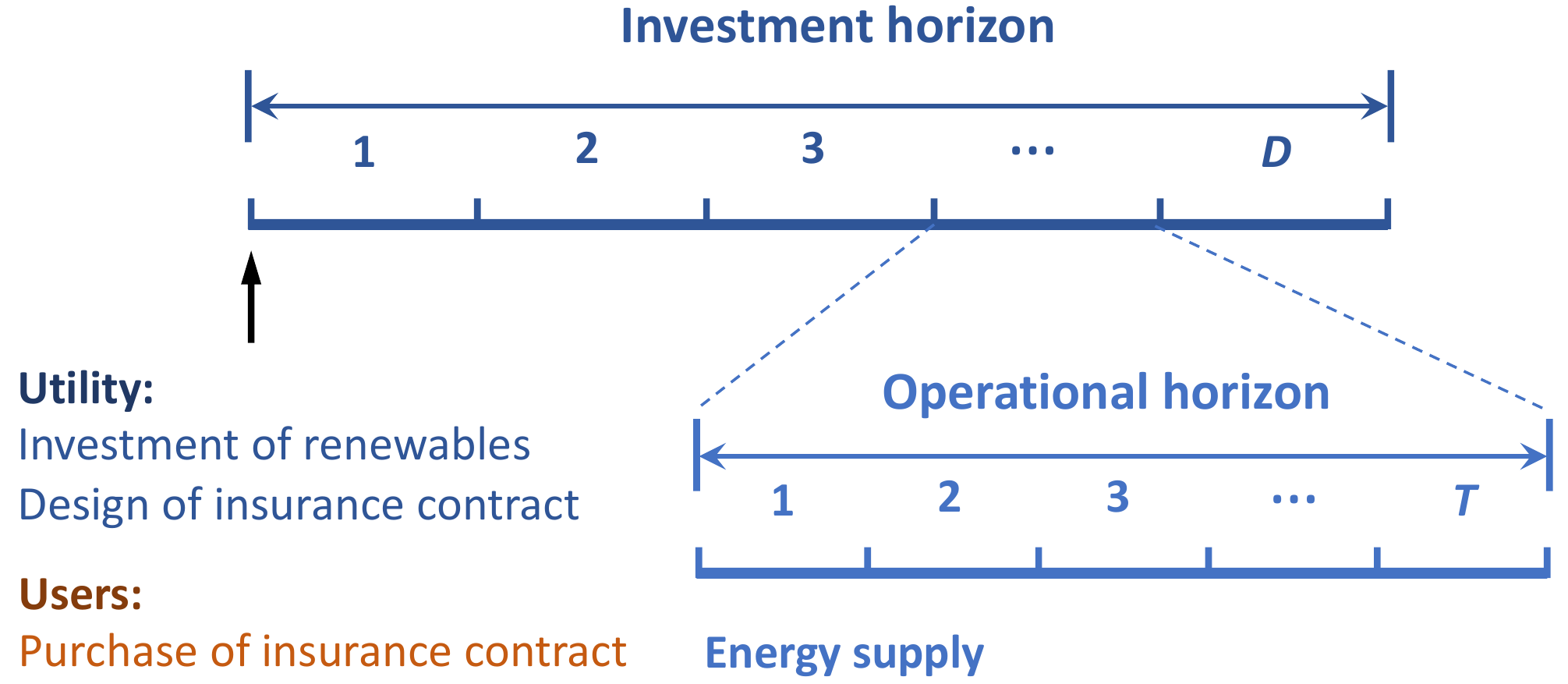}
	\vspace{-1mm}
	\caption{\small Two timescales of decision-making.}
	\label{fig:time}
	\vspace{-2.5ex}
\end{figure}

Figure \ref{fig:time} illustrates the decision-making of the utility and users in two different timescales. At the beginning of an investment horizon of $D$ days (e.g., $D$ corresponds to a few years depending on the lifespan of  renewables), the utility decides the invested renewable energy capacity and announces the insurance contract to users. Users decide whether to purchase the insurance and which type of contract items to purchase. The investment horizon is divided into operational horizons. Each operational horizon corresponds to one day, which is  divided into a set of  $T$  periods (e.g., 24 hours) $\mathcal{T}=\{1,2,3,...,T\}$. On each day, the utility supplies users with renewable energy.
Users without insurance just pay the electricity bill, but will not get reimbursed if his demand is not fully satisfied. In contrast, users who have purchased insurance will get reimbursed for the unserved demand (i.e., lost load).

\subsection{Users}

For simplicity, we assume that each user $i$'s demand $D_i^t$ in time period $t\in\mathcal{T}$ is the same across different days.\footnote{We can also model the stochastic nature of the  demand. This will make the system model more realistic, although it does not change the main analysis and insights of this paper. The reason is that what matters is the difference between the renewable energy generation and the user's total demand in each time period, which is already stochastic in our current model. We will leave such an extended study in the journal version.}  If a user has demand unserved by the utility, the lost load will incur a cost  to users due to service interruption. We denote user $i$'s private value of lost load (VOLL) at time $t$ as $V_i^t$. If user $i$'s unserved demand is $x_i^t$,  his cost of lost load at time $t$ is $V_i^t x_i^t$.

We classify users into a set  $\mathcal{K}=\{1,2,\ldots,K\}$ of types   based on the values of lost load, i.e., user $i$ belongs to type $k\in \mathcal{K}$ if his VOLL is $\bm{V}_k=(V_k^t,t\in \mathcal{T})$. The utility does not know each user's VOLL. However, we assume that utility knows each user's demand and the aggregate demand of each type, i.e., $\bm{D}_k=(D_k^t,t\in \mathcal{T})$ for any type $k\in \mathcal{K}$. Such  demand information can be obtained from smart meters, and the aggregate demand of each type can be estimated in reliability surveys \cite{schroder2015value}. We assume that ${V}_1^t\leq {V}_2^t\leq \ldots \leq {V}_K^t$  for each hour $t$, i.e., Type-1 has the lowest value.

\vspace{-1mm}
\subsection{Utility}
We consider the challenging scenario  where the utility invests in only renewable energy to supply electricity to users due to the zero-carbon target. 

The daily energy supply from the utility is uncertain due to the variability of renewable energy. If the utility invests in a total capacity $r\geq0$ of renewable energy, the daily random generation at $t\in\mathcal{T}$ as $r \Theta^{t} $, where $\Theta^{t}$ is a random variable with support $ [0,1]$. Thus, the total random electricity supplied to users at time $t$ is	$s^{t}(r,\Theta^t)=\min(D_a^t,r\Theta^t)$ where $D_a^t=\sum_{i\in \mathcal{I}}D_i^t$ is the total demand.
When the energy generation cannot satisfy users' total demand, i.e., $D_a^t>r\Theta^t$, the utility will allocate the energy according to the following rule:\footnote{This rule can be easily implemented in  expectation, e.g., the utility can shed users' load in a round-robin fashion. We leave as future work a more comprehensive study of power allocation considering network constraints. }

\vspace{-1mm}
\begin{ass} [Energy allocation rule]	
Under the capacity of renewable energy $r$, in case of insufficient energy generation, the utility allocates energy to users proportionally to the users' demands, i.e., the energy $d_i^t$ allocated to user $i$  is given by
\vspace{-1mm}
	\begin{align}
		&d_i^t(r,\Theta^t)=\frac{D_i^t}{D_a^t}\cdot  s^t(r,\Theta^t).
	\end{align}
\end{ass}
\vspace{-0.5mm}

\noindent Note that $s^{t}(r,\Theta^t)$ and $d_i^t(r,\Theta^t)$ are functions of $(r,\Theta^t)$. In the rest of paper, we sometimes omit the dependency of $s^t$ and $d_i^t$ on $(r, \Theta^t)$ for ease of exposition. However, they  should always be considered as functions of $(r, \Theta^t)$.

We assume that the utility charges a flat electricity-bill price $p$ for the actual energy consumption of each user. Since the focus of this study is insurance contract design, we will fix the value of $p$ and  assume that $p\leq {V}_1^t, \forall t\in\mathcal{T}$. Otherwise, some users with low valuation may be incentivized to curtail the load instead of paying the electricity bill.
Next, we will introduce the insurance contract design in more details.

\section{Insurance Contract Design}\label{sec:contract}

Since the utility does not know each user's private information of VOLL, we adopt contract theory to elicit users' VOLL. Specifically, the utility designs a set of contract items for all user types, such that each user will voluntarily choose the item corresponding to his own type. We will first introduce the premium and reimbursement fees in the contract. Then, we formulate the  user's problem  and the utility's problem under the insurance contract, respectively. 

\vspace{-1mm}
\subsection{Contract items}

The utility designs a set of contract items for premium fees, one for  each type (of users) in $\mathcal{K}$. The premium fee for each type $k \in \mathcal{K}$ is  $\bm{\pi}_k=(\pi_k^t,t\in \mathcal{T})$ per unit  demand paid by users. If a user chooses the type-$k$ contract and pays $\bm{\pi}_k$, the utility is obligated to compensate the user with a unit reimbursement fee for any lost demand. Note that for the demand satisfied, the user needs to pay at the electricity-bill price $p$. For the demand lost, user $i$ has a cost $V_i^t$ for the unit lost load.  Since the user will not pay the electricity bill for the lost load, we let  user $i$ of type $k$ be compensated at  $\bm{V}_k-\bm{p}=(V_k^t-p, t\in \mathcal{T})$. 

\vspace{-1mm}
\subsection{Users' problem in choosing insurance contract}

A user can decide whether to purchase the insurance. If he does, he needs to choose which type of contract to purchase.

If user $i$ chooses the contract item of type $o_i  \in \mathcal{K}$, i.e., premium fee $\bm{\pi}_{o_i}$ and reimbursement $\bm{V}_{o_i}-\bm{p}$, his net cost (scaled to one day) $CU_i({o}_i)$ is
\vspace{-0.5mm}
\begin{align}
&\sum_{t\in\mathcal{T}} {\pi_{o_i}^t}\cdot D_i^t+ \sum_{t\in\mathcal{T}} p \cdot \mathbb{E}_{\Theta^t}[d_i^t]+ \sum_{t\in\mathcal{T}} \mathbb{E}_{\Theta^t}\left[{V_i^t\cdot(D_i^t-d_i^{t})}\right]\notag\\
&- \sum_{t\in\mathcal{T}} \mathbb{E}_{\Theta^t}\left[{(V_{o_i}^t-p)}\cdot( D_i^t-d_i^{t})\right]. \label{eq:user}
\end{align}
There are  four terms in the cost formulation \eqref{eq:user}. The first term is the cost of premium fee for the type-$o_i$ contract; the second term is the electricity bill under the allocated energy amount $d_i^t$; the third part is the cost of lost load under its VOLL $V_i^t$; and the fourth term is the reimbursement from the utility. Note that if user $i$ of type $k$ chooses his own type of contract, his cost is simplified to
\vspace{-0.5mm}
\begin{align}
	CU_i(k)	:=&\sum_{t\in\mathcal{T}} {\pi_{k}^t}\cdot D_i^t+ \sum_{t\in\mathcal{T}} p \cdot D_i^t, \label{eq:owncost}
\end{align}
which only includes the premium fee and electricity bill under  the original demand. The cost of lost load is exactly reimbursed by the utility. However, if a user does not choose his own type, he will get reimbursed with a higher or lower compensation rate than his actual VOLL. Our goal is to design the contract so that each user will voluntarily choose the contract item designed for his type.

If user $i$ does not purchase insurance,  his cost  $CU_i(0)$ is
\vspace{-0.5mm}
\begin{align}
	CU_i(0)	:=& \sum_{t\in\mathcal{T}} p \cdot \mathbb{E}_{\Theta^t}[d_i^t]
	+ \sum_{t\in\mathcal{T}} \mathbb{E}_{\Theta^t}\left[{V_i^t\cdot(D_i^t-d_i^{t})}\right],
\end{align}
which only includes the costs of  electricity bill and lost load.

\subsection{Utility's problem of the optimal contract design }

At the beginning of the investment horizon, the utility optimizes the invested renewable energy capacity $r$ and premium fee $\bm{\pi}$ of the contract. We first introduce the conditions to ensure that each user chooses the contract item of his own type. Then we formulate the utility's problem to optimize the premium fee and the investment of renewable energy.

\subsubsection{Validation of the contract}

The utility announces the set of contract items to  users, and users can freely choose which type. To ensure that users will participate in the insurance contract and are reimbursed based on their actual VOLL, the insurance contract needs to satisfy \textit{Individual Rationality (IR)} and \textit{Incentive Compatibility (IC)} \cite{bolton2004contract} defined as follow.

\begin{defi} [Individual Rationality]\label{defi: IR}
	The insurance contract is individual 
	rational if a user $i$ of type $k\in \mathcal{K}$  prefers purchasing the insurance item for that type rather than not purchasing any contract, i.e., $CU_i(k) \leq CU_i(0)$.
\end{defi}

\begin{defi} [Incentive Compatibility]\label{defi: IC}
	The insurance contract is incentive
	compatible if  a user  $i$  of type  $k\in \mathcal{K}$  minimizes his cost by choosing the contract  intended for his type i.e., for any $m\in \mathcal{K}$ and $m\neq k$, $CU_i(k) \leq CU_i(m)$.
\end{defi}

A contract is valid if it can satisfy both IR and IC conditions. Theorem \ref{thm} characterizes sufficient conditions to ensure that our insurance contract is valid.
\begin{thm}\label{thm}
	The contract is valid if the following sufficient IC and IR conditions are satisfied:
	\begin{align}
		\textbf{IC}:  \pi_{k}^t- \pi_{m}^t=&   {(V_k^t-V_m^t)\cdot \mathbb{E}_{\bm{\Theta}}\left[1-\frac{s^t}{D_a^t} \right]},\notag\\ &~~~~~~~~~~~~~~~~~~\forall t\in \mathcal{T},\forall m,k \in \mathcal{K}.\label{eq:IC}\\
		&\hspace{-15ex}	\textbf{IR}:   	0\leq \pi_{k}^t\leq   {(V_k^t-p)\cdot \mathbb{E}_{\bm{\Theta}}\left[1-\frac{s^t}{D_a^t} \right]}, \forall t\in\mathcal{T},\forall k \in \mathcal{K}.\label{eq:IR}
	\end{align}
The contract is always valid.
\end{thm}

In Theorem \ref{thm}, the IC constraint \eqref{eq:IC} specifies the conditions of the premium fees between any two types, which ensures that each user will not choose other types' contract items. Note that the difference between premium fees depends on  the VOLL and renewables supply. The IR constraint  \eqref{eq:IR} ensures that the premium fee for each contract item is not too high, so that users will always choose the contract. In the later analysis, we will focus on the utility's optimization problem subject to the conditions in Theorem \ref{thm}.\footnote{We present the necessary and sufficient conditions in the appendix \cite{project4append}. The sufficient conditions in \eqref{eq:IC} and \eqref{eq:IR} make the constraints independent of individual user's private information and make the contract design  tractable.}

\subsubsection{Formulation of utility's problem}
We first denote the reimbursement payment to all users  as $C_s^t(r,\Theta^t)$:
\begin{align}
	&	C_s^t(r,\Theta^t): =  \sum_{k\in \mathcal{K}}  (V_{k}^t-p)\cdot( D_k^t-d_k^t).
\end{align}

Then, based on the sufficient conditions of IC and IR in Theorem \ref{thm}, we present the profit 	$f^{\text{Ins}}(r,\bm{\pi})$ of the utility (scaled to one day) as follows.
\begin{align}
	f^{\text{Ins}}(r,\bm{\pi})&:= \sum_{k\in\mathcal{K}} \sum_{t\in\mathcal{T}}  \pi_k^t D_k^t+\sum_{t\in \mathcal{T}} p \cdot \mathbb{E}_{\Theta^t}[s^{t}]-c_r\cdot r \notag\\
	&~~- \sum_{t\in \mathcal{T}} \mathbb{E}_{{\Theta}^t}[C_s^t(r,{\Theta^t} )]. \label{eq:profit}
\end{align}
There are  four terms in the profit formulation \eqref{eq:profit}. The first term is the revenue from all the users' premium fees; the second term is the total electricity bill under the supply $s^t$ to users; the third term is the investment cost of renewables under the unit capacity cost $c_r$ (scaled into one day); and the fourth term is the  reimbursement payment to all users.

In \textbf{Problem-Ins}, we focus on an insurance-providing utility (\textbf{Ins}-utility) whose profit is regulated to protect users  \cite{jamison2012overview}. The regulated utility is required to design the insurance to minimize all users' collective energy costs (recall single user's energy cost  in \eqref{eq:owncost}), constrained on a minimum regulated profit of $\xi\geq 0$.\footnote{We also consider a profit-seeking utility that maximizes her profit in the appendix \cite{project4append}. However, we show that the insurance contract has no benefit compared with the no-insurance case if the utility is profit-seeking.}   As shown in \eqref{eq:owncost}, when a user chooses his own type of contract, his cost only includes the premium fee and the electricity bill under the original demand. Since the electricity-bill price and the original demand are fixed, we can simply minimize the total premium fees as in the objective \eqref{eq:obj} to minimize users' collective energy costs.

\textbf{Problem-Ins: Regulated utility under insurance contract}
			\begin{align}
	\min~ & \sum_{k\in\mathcal{K}}\sum_{t\in\mathcal{T}} \pi_k^t D_k^t \label{eq:obj}\\	
	\text{s.t.}~
	& f^{\text{Ins}}(r,\bm{\pi}) \geq \xi,\label{eq:profitbound}\\
	&\eqref{eq:IC}-\eqref{eq:IR},\notag\\
	\text{var}: &r\geq0, \bm{\pi},\notag
\end{align}
The constraint \eqref{eq:profitbound} regulates the utility's profit to be no greater than a lower bound $\xi\geq0$. The constraints \eqref{eq:IC} and \eqref{eq:IR} are the IC and IR conditions, respectively. We denote the optimal solution to \textbf{Problem-Ins} as $(r^*,\bm{\pi}^*)$.  \textbf{Problem-Ins} is challenging to solve due to the non-convexity brought by \eqref{eq:IC}. We will present the solution method in Section \ref{sec:solution}. Before the solution method, we present two performance benchmarks in the next section.

\section{Benchmarks}

We consider two benchmarks to compare the results under the insurance contract. One is the social-optimum case, and the other is the no-insurance case. 

\subsection{Social-optimum benchmark (SO)}  
For the \textbf{SO} benchmark,  a social planner invests in renewable energy to minimize the system cost. The system cost, denoted as $C^{so}(r)$,  consists of the investment cost of renewable energy $c_r\cdot r$ and the expected cost of lost load $ \sum_{t\in \mathcal{T}} \mathbb{E}_{\Theta^t}[C_s(r,{\Theta^t} )]$. The social-cost minimization problem is formulated as

\vspace{-0.2ex}
\textbf{Problem-SO: Social-cost minimization}
\begin{align}
	\min~C^{\text{SO}}(r):= & c_r\cdot r + \sum_{t\in \mathcal{T}} \mathbb{E}_{\Theta^t}[C_s(r,{\Theta^t} )] \\	
	\text{s.t.}~
	& r\geq 0,\\
	\text{var}: &r.\notag
\end{align}
The objective function $C^{\text{SO}}(r)$ is convex in $r$. \textbf{Problem-SO} can be easily solved by standard context optimization techniques. We let $r^\dagger$ be the optimal solution to \textbf{Problem-SO}. 

\subsection{No-insurance benchmark (NoIns)}  

For the \textbf{NoIns} benchmark, we consider a regulated utility that does not provide insurance (\textbf{NoIns}-utility). To reduce users' lost load and thus minimize users' total energy cost,  the utility maximizes the invested capacity of renewable energy under a regulated profit lower bound. The no-insurance optimization problem is formulated as

\textbf{Problem-NoIns: Regulated utility without insurance}
\begin{align}
	\max~ &r \notag\\	
	\text{s.t.}~
	& f^{\text{No}}(r) \geq \xi,\\
	\text{var:}~ &r\geq0, \notag
\end{align}
where the profit of the utility is denoted as a function $f^{No}(r)$ of the invested capacity $r$, given by
\begin{align}
	f^{\text{No}}(r)=&\sum_{t\in\mathcal{T}} p \cdot \mathbb{E}_{\Theta^t}[s^{t}]-c_r\cdot r.
\end{align}

We let $r^\ddagger$ be the optimal solution to \textbf{Problem-NoIns}. As $f^{\text{No}}(r)$ is concave, \textbf{Problem-NoIns} is a concave maximization problem. More specifically, we have	$f^{\text{No}}(0)=0$ and $f^{\text{No}}(+\infty)<0$.  Thus, if there exists $r'$, such that $f^{\text{No}}(r')\geq \xi$, the optimal solution $r^\ddagger$ to \textbf{Problem-NoIns} is achieved at $f^{\text{No}}(r^\ddagger)=\xi$ because $f^{\text{No}}(r)$ is either decreasing or first increasing and then decreasing in $r$. The equation $f^{\text{No}}(r)=\xi$ has either one root if $\max_r f^{\text{No}}(r)\leq 0$, or two roots if $\max_r f^{\text{No}}(r)>0$. If there are two roots,  we select the larger one to maximize the capacity. Such a root can be found through standard numerical methods like bisection search \cite{solanki2014role}.

\section{Analysis of insurance contract}\label{sec:solution}

\textbf{Problem-Ins} is challenging to solve due to non-convexity. With the help of the two benchmarks, we will first show how to solve  \textbf{Problem-Ins} for the \textbf{Ins}-utility, and then present analytical results regarding the performance of the insurance contract. We include all the proofs in the online appendix \cite{project4append} due to the page limit.

 \subsection{Solution method for \textbf{Ins}-utility}

\textbf{Problem-Ins} is non-convex due to the equality constraint \eqref{eq:IC}. To solve it, we first solve the optimal capacity $r^*$ by constructing a convex problem from the original problem. Then, under $r^*$, we solve the optimal premium fee $\bm{\pi}^*$ based on the binding condition of the IR constraint \eqref{eq:IR}.

Specifically, the convex problem to derive $r^*$ and the binding condition of the IR constraint \eqref{eq:IR} are affected by the electricity-bill price $p$. We present the results in Proposition \ref{prop:ins}.

\begin{prop}[Solution structure in \textbf{Problem-Ins}]\label{prop:ins}
	Assume that the profit lower-bound \eqref{eq:profitbound} is feasible. Recall that $r^\dagger$ is the optimal solution to \textbf{Problem-SO}. We let
	\begin{align}
		& \underline{L}=\frac{c_r\cdot r^\dagger+\xi}{\sum_t\mathbb{E}[s^{t\dagger}]},	\\
		&\overline{L}=\frac{\xi+c_r\cdot r^\dagger+\sum_tV_1^t(D_a^t-\mathbb{E}[s^{t\dagger}])}{\sum_tD_a^t}.
	\end{align}
	
	\begin{itemize}
		
	\item If  $\underline{L}	<p<\overline{L}$, we have
$r^*=r^\dagger$.
	
		\item if $p\leq \underline{L}$, the IR constraint \eqref{eq:IR} reaches the upper  bound for all types and hours. We have  $r^*=r^\ddagger$, which is the optimal solution to \textbf{Problem-NoIns}. 
		\item if $p\geq\overline{L}$,   IR constraint \eqref{eq:IR} of Type-1 reaches the lower bound for all hours. The optimal capacity $r^*$ is equal to the optimal solution of the following concave problem.
		\begin{align}
	\max~ &r \notag\\	
	\text{s.t.}~
	& f^{\text{No}}(r)-\sum_{t\in \mathcal{T}}{(V_1^t-p)\cdot \mathbb{E}_{\bm{\Theta}}\left[D_a^t-{s^t} \right]} \geq \xi,\\
	\text{var:}~ &r\geq0 \notag
\end{align}
	\end{itemize}
	
	Furthermore, the utility's profit always achieves the lower bound $\xi$ in all the above three cases, i.e.,  $f^{\text{Ins}}(r^*,\bm{\pi}^*) =\xi$.
\end{prop}

As Proposition \ref{prop:ins} shows, the binding condition of IR \eqref{eq:IR} is affected by the electricity-bill price. If the price is too low, the premium fees need to ensure enough revenues for the utility, so that the IR constraints reach the upper bound. If the price $p$ is too high, the premium fees should be set low to provide benefits to users, and the IR constraints reach the lower bound. 

\begin{figure}[t]
	\centering
	\includegraphics[width=2.4in]{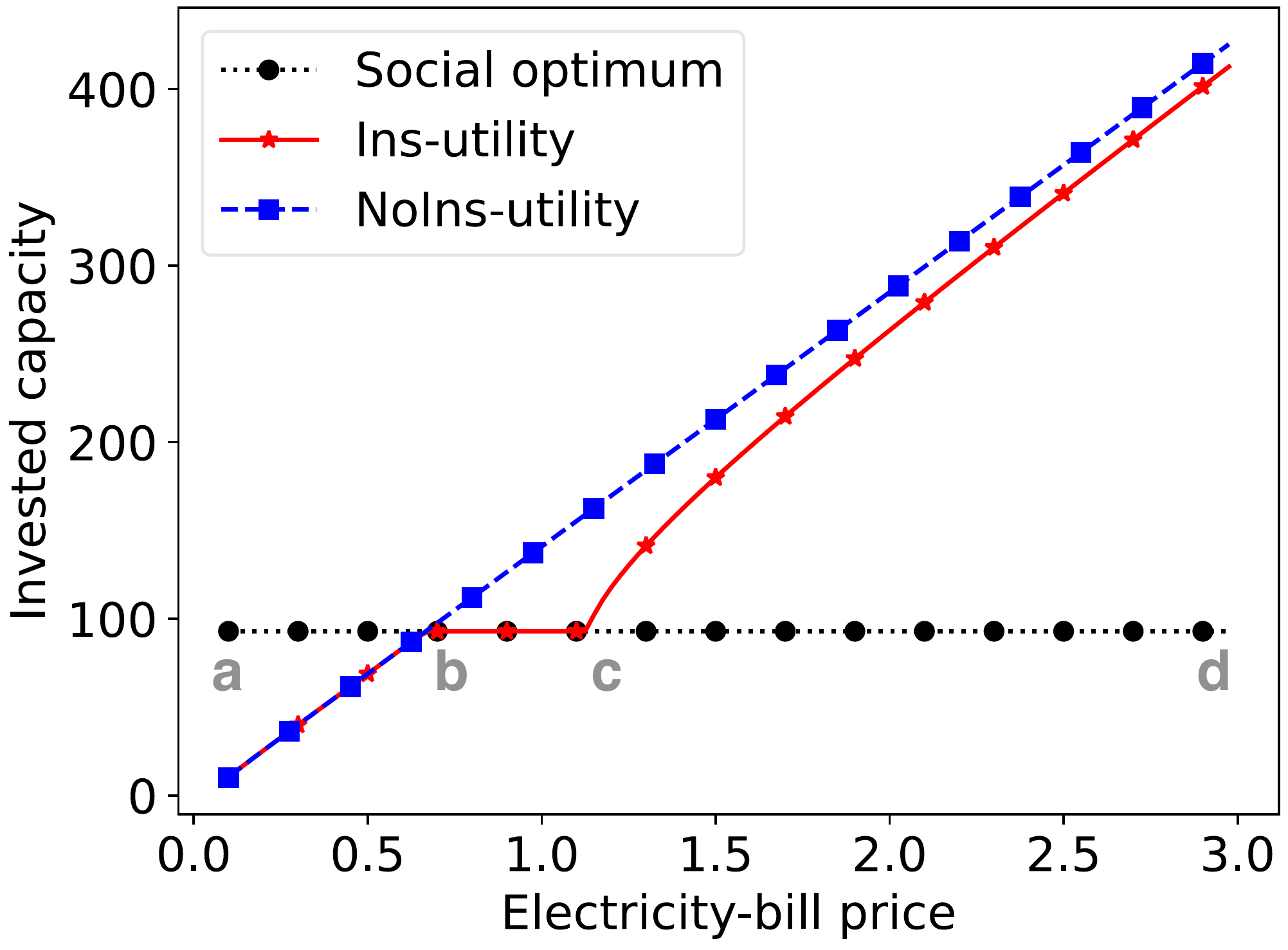}
	\vspace{-2mm}
	\caption{\small Illustration of solution structure for \textbf{Ins}-utility.}
	\label{fig:capacity}
	\vspace{-2ex}
\end{figure}

In Figure \ref{fig:capacity}, we illustrate the results in Proposition \ref{prop:ins}, as the electricity-bill price varies. We show the optimal invested capacity  under the benchmark \textbf{SO} (black curve), \textbf{NoIns}-utility (blue curve), and \textbf{Ins}-utility  (red curve), with respect to the electricity-bill price. The red curve (\textbf{Ins}-utility) consists of three parts.  Within the lowest-price range (a-b), the investment level overlaps with that of the \textbf{NoIns}-utility. Within the medium price range (b-c),  the investment level reaches that of the social optimum (\textbf{SO}). Within the highest-price range (c-d), the investment level coincides with neither  \textbf{NoIns}-utility nor the social optimum (\textbf{SO}).

Based on Proposition \ref{prop:ins}, it is straightforward to first  solve $r^*$ by the equivalent convex problems. Then, under the solved $r^*$, we can  compute the optimum premium fee $\bm{\pi}^*$ based on the IC constraint \eqref{eq:IC} and the IR constraint \eqref{eq:IR}, which we provide a sketch in the following.

\begin{itemize}
	\item   If $p\leq \overline{L}$,  we compute $\bm{\pi}^*$ directly using  the upper bound of \eqref{eq:IR} for all types and hours.
	\item  If $\underline{L}	<p<\overline{L}$, we derive $\bm{\pi}^*$ using the solution to the set of linear  equations $f^{\text{Ins}}(r^*,\bm{\pi})=\xi$ and  \eqref{eq:IC}. 
	
	\item  If $p\geq \overline{L}$, we compute $\bm{\pi}^*$  using \eqref{eq:IC} and the lower bound of \eqref{eq:IR} constraint for all types and hours.
	
\end{itemize}

\subsection{Benefits of insurance contract}
Compared with \textbf{Problem-NoIns}, \textbf{Problem-Ins} will consider the VOLL of users in the insurance design and renewable planning, which improves social welfare and benefits the users.
Next, we present two propositions to demonstrate the benefits of the \textbf{Ins}-utility compared with the \textbf{NoIns}-utility.

\begin{prop} \label{prop-2}
	(Social cost reduction) The optimal social cost under \textbf{Ins}-utility is no larger than that under  \textbf{NoIns}-utility.
\end{prop}

\begin{prop} \label{prop-3}
	(Reduction in users' energy cost) All users' total cost under \textbf{InsR}-utility is no larger than that under \textbf{NoIns}-utility.
\end{prop}

Proposition \ref{prop-2} shows that the insurance contract will not increase the social cost compared with the case that the utility does not provide insurance. Proposition \ref{prop-3} shows that  users' collective cost is always no larger under \textbf{Ins}-utility than that under \textbf{NoIns}-utility.

Although individual users may not always have lower costs under \textbf{Ins}-utility than under \textbf{NoIns}-utility, they can still see a significant cost reduction if the electricity-bill price is medium and the heterogeneity of types is at a low level. We will demonstrate this in simulations of Section \ref{sec:sim}.

Finally, since the IR constraint is satisfied in the insurance, under \textbf{InsR}-utility, a user's energy cost when purchasing insurance is always no larger compared to not purchasing.

\section{Numerical  studies}\label{sec:sim}

We conduct extensive numerical studies to show how the insurance contract can improve social welfare and reduce users' costs. In particular, we  look into the impact of several system parameters, including  the  electricity-bill price, heterogeneity of types, and the uncertainty of renewable energy.

\subsection{Setup}

As an illustrative example, we consider four users\footnote{The utility makes decisions based on the aggregate demand of each type. Thus, the simulation results in this section can still hold if we fix the aggregate demand and consider multiple users.} of four types  with distinct values of lost load  $V_1^t\leq V_2^t \leq V_3^t\leq  V_4^t $ to capture the key feature of type heterogeneity. The VOLL for each user is the same for all the time slots $t\in \mathcal{T}$. We consider the identical demand $D_k^t=D$ for each type $k\in \mathcal{K}$ at each time slot $t\in \mathcal{T}$. We simulate  both uniform distribution and truncated normal distribution within $[0,1]$ for the output factor  $\Theta^t$.   We let the factor  $\Theta^t$ be independent and identically distributed for all the time slots $t\in \mathcal{T}$. To examine the cost reduction  under insurance, what matters are relative values between the capital cost $c_r$, electricity-bill price $p$, and VOLLs. Thus, we set the capital cost at $c_r=1$, and scale the parameters $p$ and VOLLs accordingly.\footnote{The range of practical electricity-bill prices and capital costs of renewables can be quite wide. The US electricity-bill price is about {0.1-0.3 \$/kWh}\cite{electricityrate}. The capacity cost of PV solar and onshore wind in 2020 can be about 800-2000\$/kWh for 20 years \cite{renewstatus2021}, i.e., about {0.1-0.3\$/kWh} per day. Thus, if we scale the renewable-energy capacity cost to $c_r=1$, the electricity-bill price $p$ can be in the range about 0.33-3.}
\begin{figure*}[t]
	\centering
	\hspace{-3ex}
	\subfigure[]{
		\raisebox{-2mm}{\includegraphics[width=2.1in]{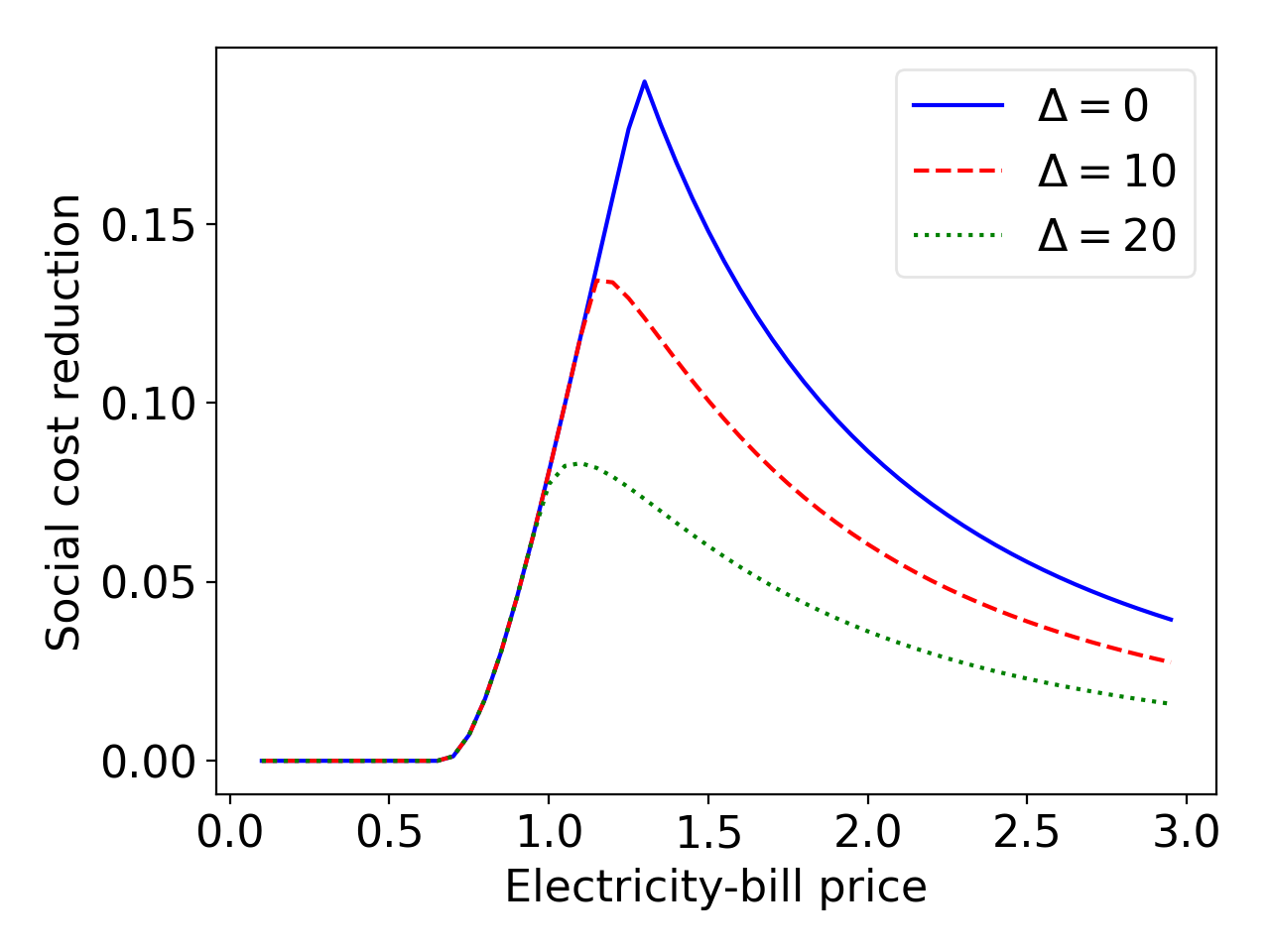}}}
	\hspace{-3ex}
	\subfigure[]{
		\raisebox{-2ex}{\includegraphics[width=2.1in]{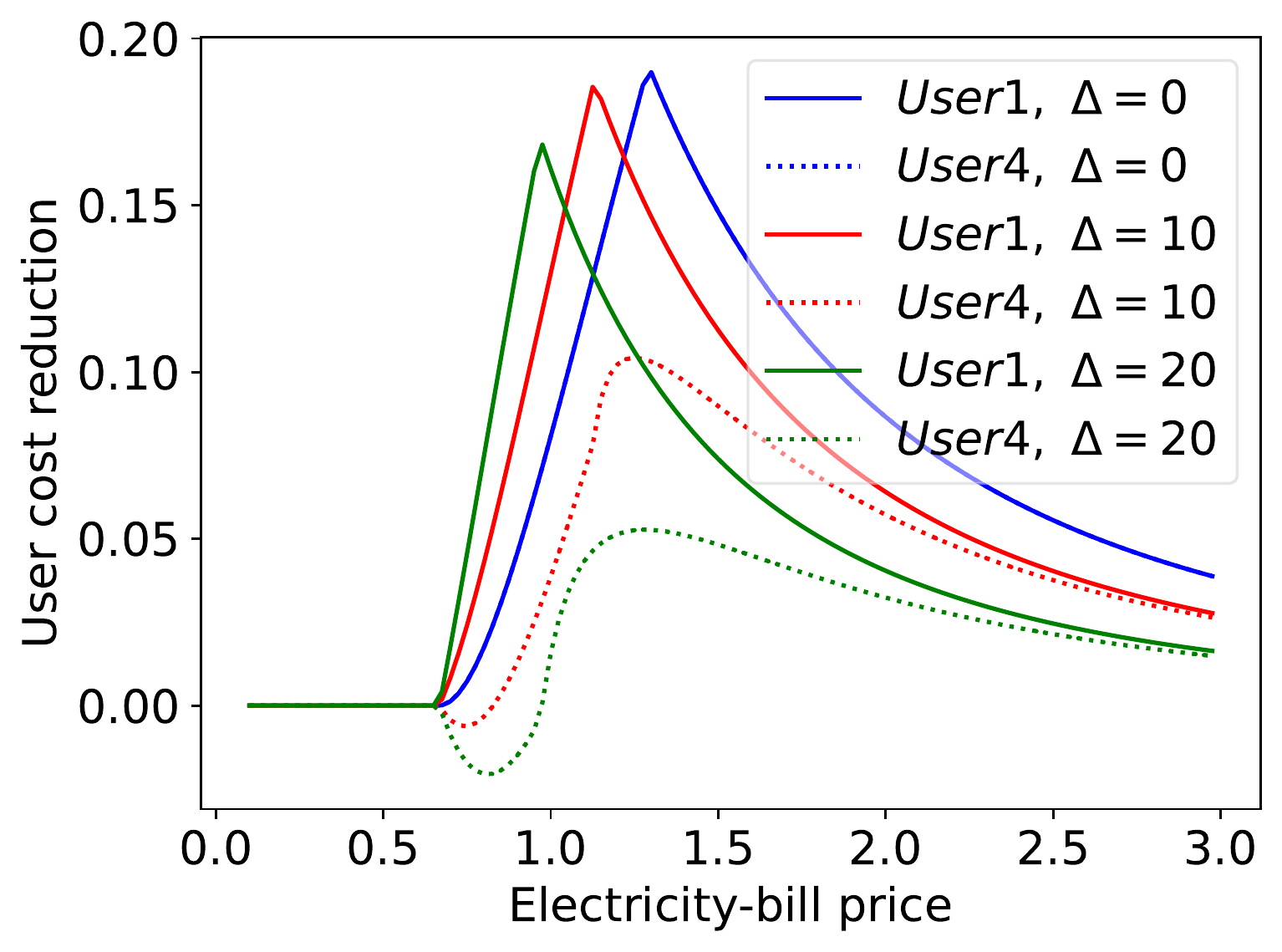}}}
	\subfigure[]{
		\raisebox{-2ex}{\includegraphics[width=2.1in]{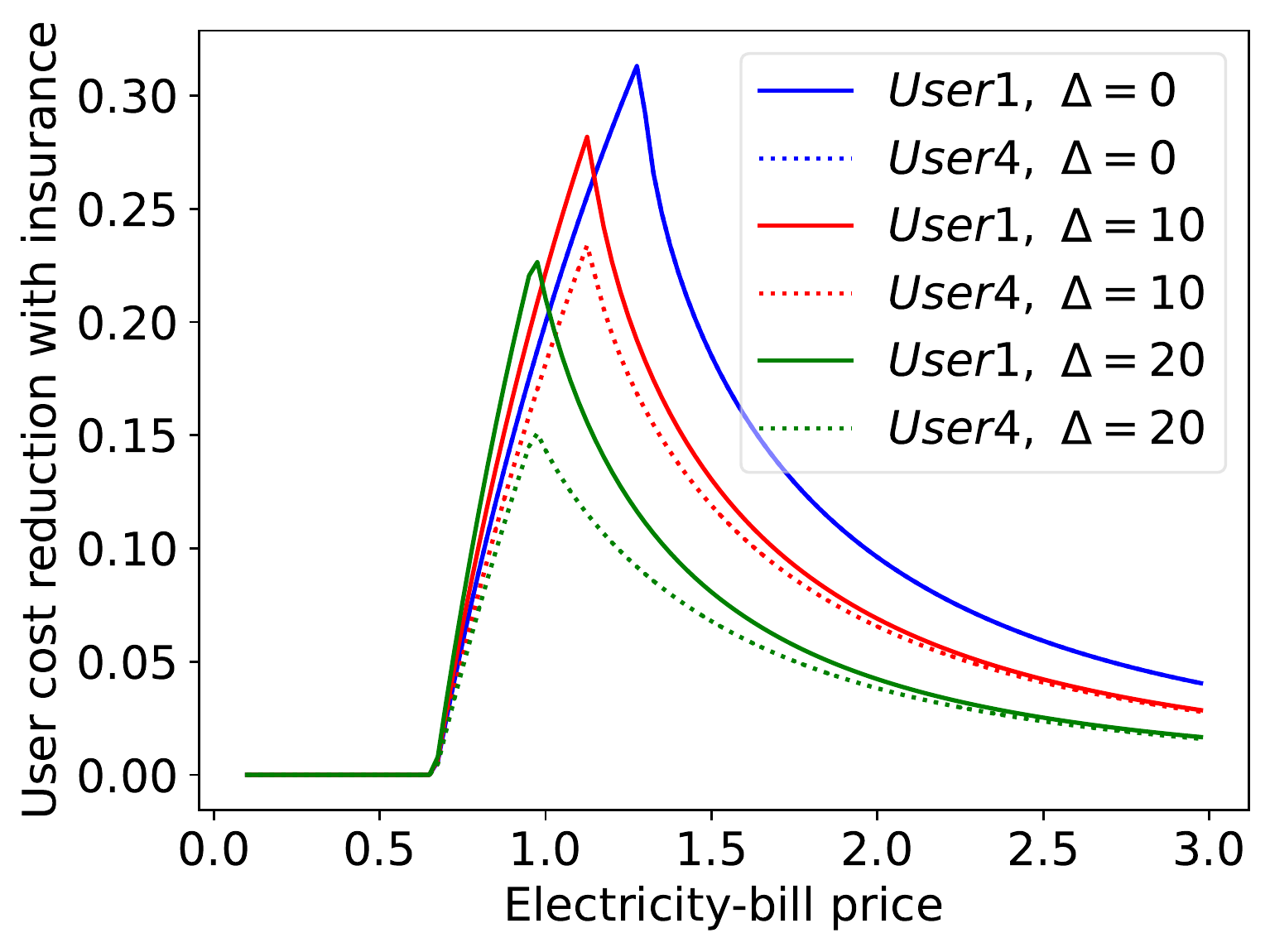}}}
	\vspace{-2mm}
	\caption{\small (a) Social-cost reduction under  \textbf{Ins}-utility compared with \textbf{NoIns}-utility;	(b) Users' cost reduction under \textbf{Ins}-utility compared with \textbf{NoIns}-utility; (c) Under \textbf{Ins}-utility, users' cost reduction by purchasing insurance compared with not purchasing.}
	\label{fig:benefit}
	\vspace{-4mm}
\end{figure*}

\begin{figure*}[t]
	\centering
	\hspace{-3ex}
	\subfigure[]{
		\raisebox{-2mm}{\includegraphics[width=2.1in]{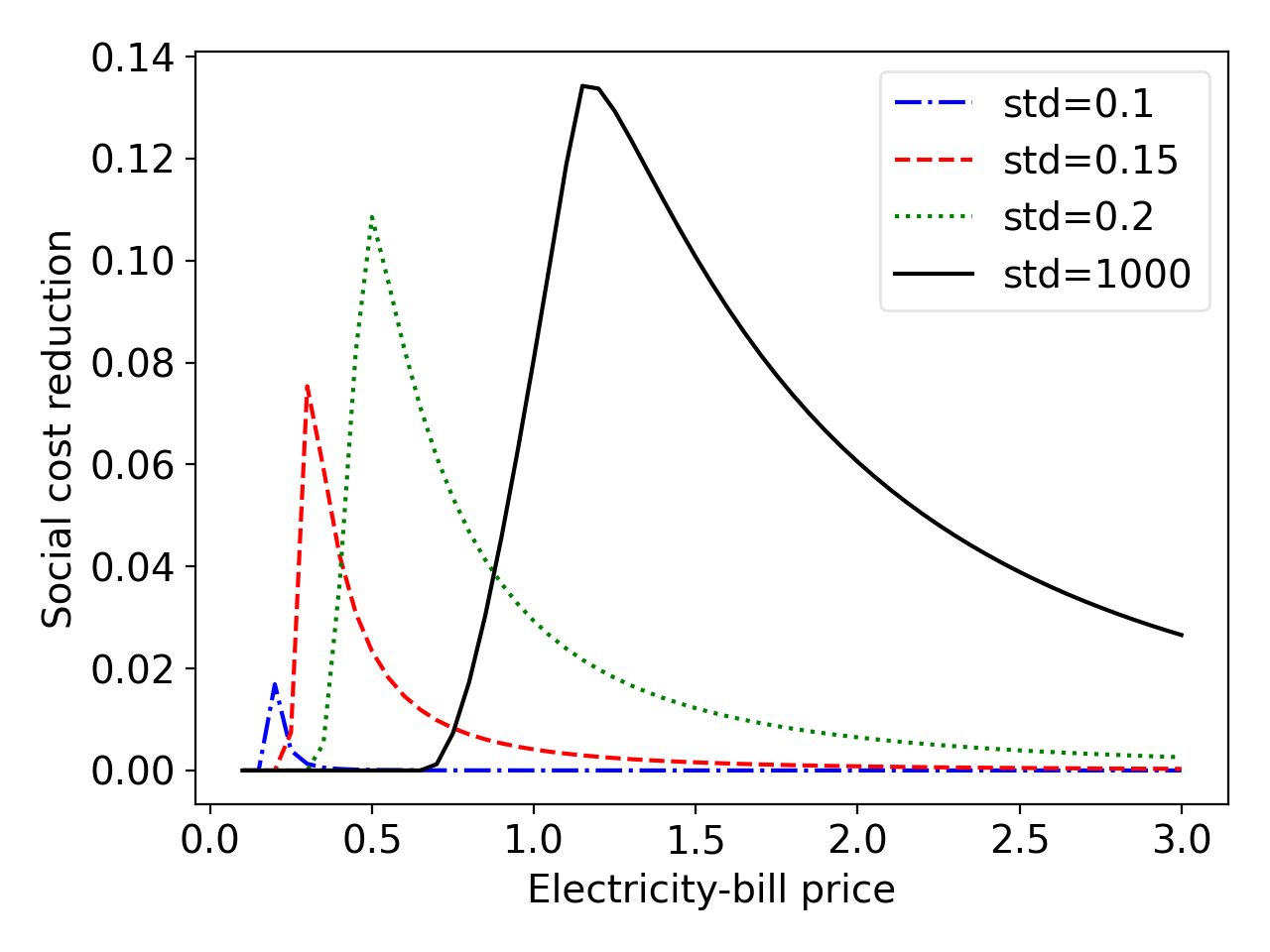}}}
	\hspace{-3ex}
	\subfigure[]{
		\raisebox{-2ex}{\includegraphics[width=2.1in]{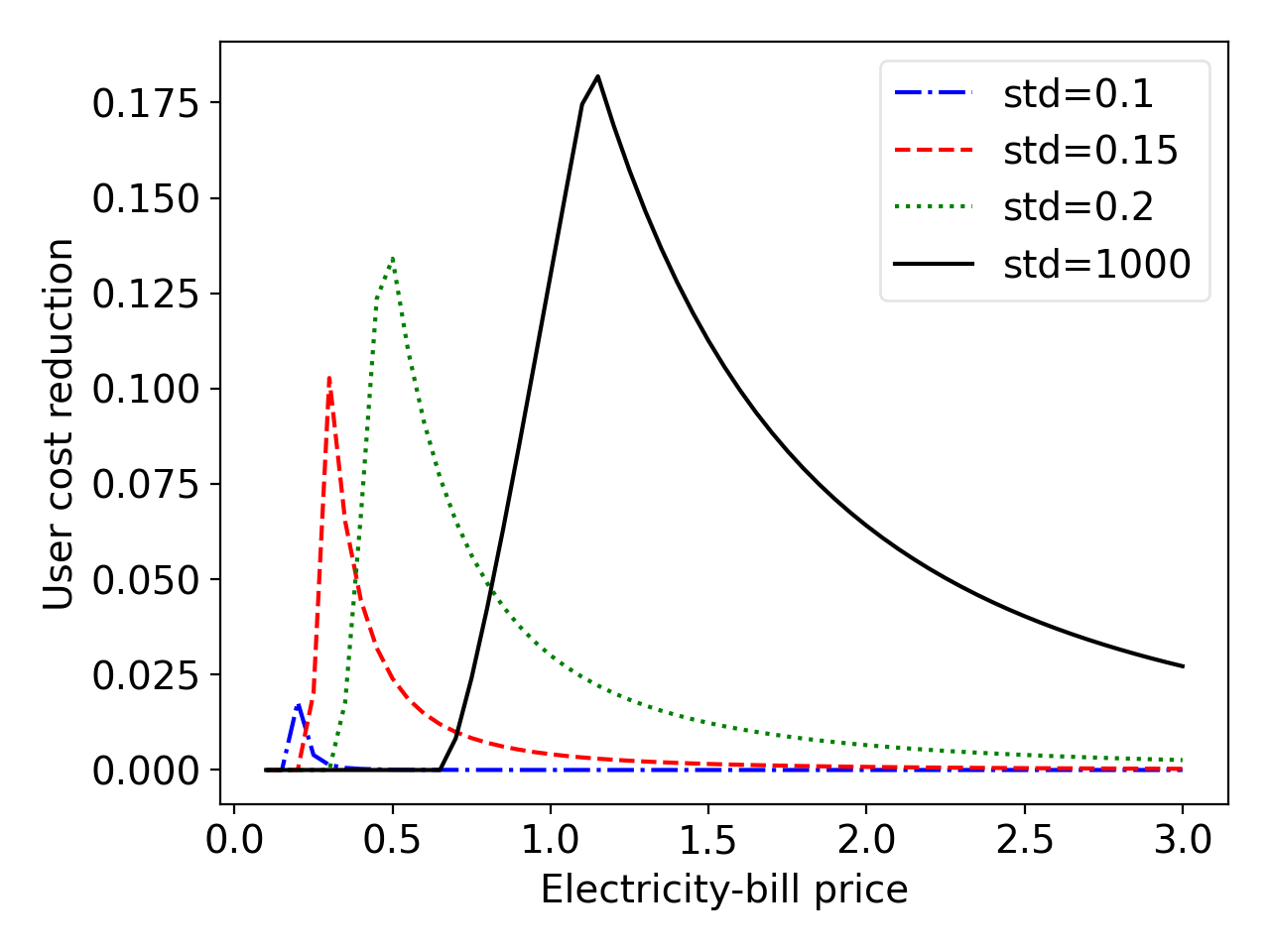}}}
	\subfigure[]{
		\raisebox{-2ex}{\includegraphics[width=2.1in]{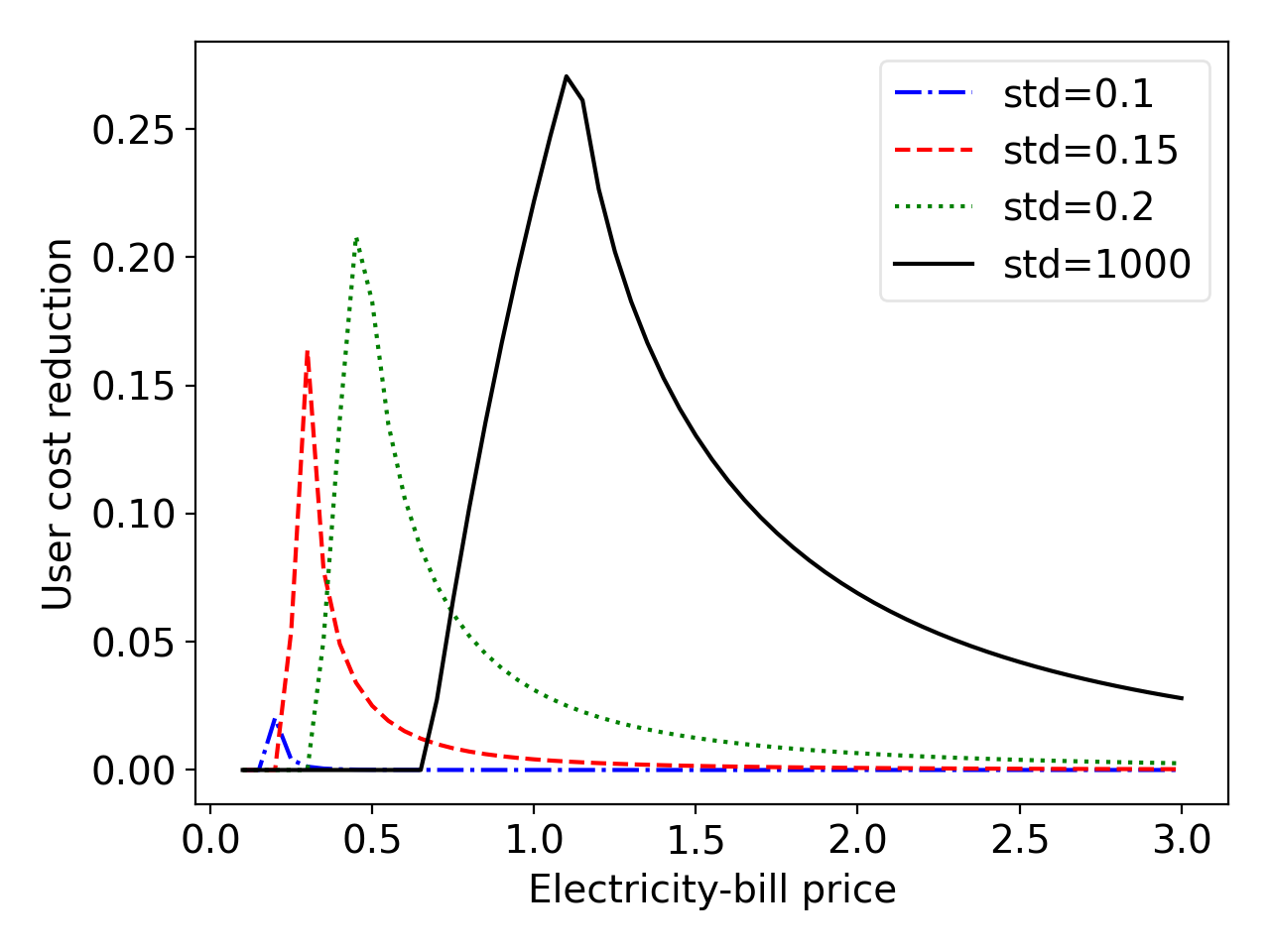}}}
	\vspace{-2mm}
	\caption{\small (a) Social cost reduction under \textbf{Ins}-utility comapred with  \textbf{NoIns}-utility;	(b)  Type-1 user's cost reduction under  \textbf{Ins}-utility comapred with \textbf{NoIns}-utility; (c) Under \textbf{Ins}-utility, Type-1 user'scost reduction by purchasing insurance compared with not purchasing.}
	\label{fig:uncertainty}
	\vspace{-4mm}
\end{figure*}

 In Figures \ref{fig:benefit} and Figures \ref{fig:uncertainty}, the horizontal axis denotes the electricity-bill price $p$.  In Figures \ref{fig:benefit}, we fix the uniform distribution of renewable energy and compare different VOLLs of users.  We let  $V_1+V_4=40$, $V_4-V_1=\Delta$. Each curve corresponds to a different value of  $\Delta$. In Figure \ref{fig:uncertainty}, we fix the VOLLs of users at $V_1=V_2=15$ and $V_3=V_4=25$, and compare different levels of uncertainty of the truncated normal distribution.  Each curve corresponds to a different value of the standard deviation (std).

\subsection{Benefits of insurance contract}

\subsubsection{Social cost reduction}

Comparing the \textbf{Ins-utility} with \textbf{NoIns}-utility, adopting the optimal insurance can significantly reduce the  social cost  when the electricity bill price is moderate.  As shown in Figure  \ref{fig:benefit}(a), the social cost reduction can be  nearly 20\% at $\Delta=0$ and about 14\% at $\Delta=10$ when the electricity price is moderate (around $1<p<1.5$).

\subsubsection{Users' cost reduction}

Comparing the \textbf{Ins}-utility with \textbf{NoIns}-utility, adopting the optimal insurance can significantly reduce the users' cost. As shown in Figure  \ref{fig:benefit}(b), Type-1 user's cost reduction can be up to 17\%$ \sim$19\% when the electricity price is moderate (around $1<p<1.5$).  Type-4 user's cost can be reduced by over 10\% if the difference $\Delta$ is not too high (e.g., $\Delta=0$ and $10$). 

Furthermore, under \textbf{Ins}-utility, users' costs never increase if they buy insurance compared with not buying insurance. As shown in Figure  \ref{fig:benefit}(c), by purchasing insurance, Type-1 user's cost reduction can be up to 23\%$ \sim$32\% when the electricity price is moderate. Type-4 user's cost reduction can be  about 15\%$ \sim$32\% (around $1<p<1.5$).

\subsection{Impact of the electricity-bill price, the type heterogeneity, and the renewable uncertainty}

As we elaborate below, our simulation results  show that the largest benefit of the optimal insurance contract (in terms of increasing social welfare and reducing users' costs) is achieved at a medium electricity-bill price together with a low level of type heterogeneity and a high level of renewable uncertainty.

\subsubsection{Electricity-bill price} The maximum cost reduction is achieved at a medium electricity-bill price (e.g., $1<p<1.5$ in Figures \ref{fig:benefit}). When the price is too high (e.g., $p>2.5$ in Figures \ref{fig:benefit}) or too low (e.g., $p<0.6$ in Figure \ref{fig:benefit}), the cost reduction becomes close to zero. The reason is that  too high (or too low) a price will lead to large (or zero) investment of renewables, reducing the benefit of insurance.

\subsubsection{Type heterogeneity} In Figure \ref{fig:benefit}(a)-(c), the blue curves correspond to the identical VOLL of two users with $\Delta=0$,  the red curves correspond to $\Delta=10$, and the green curves correspond to  $\Delta=20$. A higher  $\Delta$ implies a higher heterogeneity. In all three subfigures,   as the difference $\Delta$ increases, the benefit of the insurance, including the social cost reduction and users' cost reduction, will go down. The reason is that the utility needs to satisfy the IC constraint \eqref{eq:IC} in the insurance design. A higher heterogeneity will incur an additional cost due to this IC constraint. In contrast, if all the types are identical,  the IC constraint is trivially satisfied.

\subsubsection{Renewable uncertainty} In Figures \ref{fig:uncertainty}, a higher variance of the truncated normal distribution will increase the maximum cost reduction. For example, when the variance of renewable energy is close to zero (e.g., $\text{std}=0.1$), the maximum social-cost reduction in Figure \ref{fig:uncertainty}(a) is close to 0 (below 2\%). The higher uncertainty makes the insurance more important.

\section{Conclusion}
\vspace{-1ex}
This work proposes the optimal insurance contract design between the utility and end-users, which can hedge end-users' risk and incorporate their heterogeneous reliability preferences in the utility's planning of renewable energy. The joint optimization of contract design and renewable energy planning involves solving a challenging non-convex optimization problem.  We resolve this difficulty by revealing important structural properties of the optimal solution, using the help from two benchmark problems: the no-insurance benchmark and the social-optimum benchmark.   Compared with the no-insurance benchmark,  we demonstrate that the optimal insurance contract can improve social welfare and reduce users' cost.  However, the specific performance gain is affected by a number of factors, such as the electricity-bill price, the heterogeneity of types, and the uncertainty of renewables. Our future work will model the random demand and incorporate energy storage.

\bibliographystyle{IEEEtran}
\bibliography{storage}

\newpage
\appendix
\section*{Appendix I. Proof of Theorem \ref{thm}}
 We characterize the sufficient and necessary conditions for incentive compatibility and individual rationality, respectively. Then, we  provide the sufficient conditions.
\subsection{Incentive compatibility} Considering any user $i$ of type $k$, the condition of IC requires	$CU_i(o_i^{k}) \leq CU_i(o_i^m),\forall m\in \mathcal{K}$, which  leads to 
 			\begin{align}
&\sum_t {{\pi_{k}^t}}\cdot D_i^t\leq \sum_t {{\pi_{m}^t}}\cdot D_i^t \notag\\
&+\sum_t\mathbb{E}\left[{V_k^t\cdot(D_i^t-d_i^{t})}-V_{m}^t\cdot( D_i^t-d_i^{t})\right], \forall m\in \mathcal{K}. \label{eq:nsi}
 	\end{align}
The inequality set above is the sufficient and necessary condition of IC for user $i$ of type $k$. Considering all the users, these inequality sets make up the sufficient and necessary conditions of IC for the contract design.

 
 
 
One sufficient condition that makes \eqref{eq:nsi}  feasible for all the users is 
 	\begin{align}
&\pi_{k}^t- \pi_{m}^t= \mathbb{E}\left[{V_k^t\cdot(1-\frac{s^t}{ D_a^t})}-{V_{m}^t}\cdot(1-\frac{s^t}{D_a^t})\right],\notag \\
&~~~~~~~~~~~~~~~~~\forall t\in \mathcal{T}, \forall m,k \in \mathcal{K},\label{eq:ICa}
 	\end{align}
which makes  \eqref{eq:nsi} always achieve at the equality. This sufficient condition makes the constraint of IC independent of individual user's information, which only depends on the type information. Under this condition, the costs of any user choosing any types will be the same. Furthermore,  \eqref{eq:ICa} is feasible.

\subsection{Individual rationality}

Considering  any user $i$ of type $k$, we have $CU_i(o_i^{k}) \leq CU_i(o_i^o)$ for individual rationality, which leads to
 \begin{align}
\sum_t {{\pi_{k}^t}}D_i^t\leq \sum_t{\mathbb{E}\left[(V_k^t-p)\cdot (1-\frac{s^t}{ D_a^t} )D_i^t\right]}. 
 \end{align}
The inequality set above is the sufficient and necessary condition of IR for user $i$ of type $k$. Considering all the users, these inequality sets make up the sufficient and necessary conditions of  IR for the contract design.

Considering non-negative premium fees, we have one sufficient condition  is
  	 		\begin{align}
0\leq \pi_{k}^t\leq   {(V_k^t-p)\cdot \mathbb{E}[(1-\frac{s^t}{D_a^t} )]},\forall t\in \mathcal{T}, \forall k \in \mathcal{K}, \label{eq:IRa}
 \end{align} which also
  makes the constraint of IR independent of individual user's information.

To sum up, we have Theorem 1 proved. \qed

\section*{Appendix II. Proof of Proposition \ref{prop:ins}}

We will first prove that the optimal solution is achieved at the lower bound of \eqref{eq:profitbound} and then we prove the solution structures under the three electricity-bill price conditions respectively.

To begin with, we can rewrite the IC constraint \eqref{eq:IC} and IR constraint \eqref{eq:IR} of \textbf{Problem-Ins} based on the Type 1's variable $\bm{\pi}_1$ into  \eqref{eq:ICN} and \eqref{eq:IRN} as follows.
	\begin{align}
&  \pi_{k}^t= \pi_{1}^t+  {(V_k^t-V_1^t)\cdot \mathbb{E}_{\bm{\Theta}}\left[1-\frac{s^t}{D_a^t} \right]},\forall t\in \mathcal{T},\forall k \in \mathcal{K}.\label{eq:ICN}\\
		&  	0\leq \pi_{1}^t\leq   {(V_1^t-p)\cdot \mathbb{E}_{\bm{\Theta}}\left[1-\frac{s^t}{D_a^t} \right]}, \forall t\in\mathcal{T},\forall k \in \mathcal{K}.\label{eq:IRN}
	\end{align}
	
Based on 	\eqref{eq:ICN}, we have
	
	\begin{align}
&  \sum_t\sum_ k \pi_{k}^t  D_k^t= \sum_t  \left(\pi_{1}^tD_a^t+  \sum_kD_k^t{(V_k^t-V_1^t)\cdot \mathbb{E}\left[1-\frac{s^t}{D_a^t} \right]}\right ). \label{eq:type1re}
	\end{align}

Based on \eqref{eq:IRN} and \eqref{eq:type1re}, we have
		\begin{align}
&  f^l(r)\triangleq \sum_t  \left( \sum_kD_k^t{(V_k^t-V_1^t)\cdot \mathbb{E}\left[1-\frac{s^t}{D_a^t} \right]}\right )\notag \\&\leq \sum_t\sum_ k \pi_{k}^t  D_k^t\notag\\
&\leq \sum_t  \left(  \sum_kD_k^t{(V_k^t-p)\cdot \mathbb{E}\left[1-\frac{s^t}{D_a^t} \right]}\right ) \triangleq  f^u(r). \label{eq:threshold}
	\end{align}

\subsection*{A. The optimal solution is achieved at the lower bound of \eqref{eq:profitbound}}

We prove that the optimal solution is achieved at the lower bound of \eqref{eq:profitbound} by contradiction. Suppose that the optimal solution is only achieved with  $f^{\text{Ins}}({r}^*,\bm{\pi}^*)> \xi$.
\begin{itemize}
    \item If \eqref{eq:IRN} is not lower-bounded for some hour $t$, we can always reduce $\pi_1^{t*}$ a bit to $\pi_1^{t'}$  such that $f^{\text{Ins}}({r}^*,\bm{\pi}')> \xi$. This reduces the objective, which contradicts the optimal solution.
    \item If \eqref{eq:IRN} is lower-bounded for all the hours, 
we can always find $r'>{r}^*$ such that $f^{\text{Ins}}({r}',\bm{\pi}^*)= \xi$ because $f^{\text{Ins}}({r},\bm{\pi})\rightarrow -\infty$ as $r\rightarrow +\infty$.  Based on \eqref{eq:ICN}, since $s$ is non-decreasing in $r$, the total premium fee, i.e., the objective will be not be increased, which contradicts  that the optimal solution is only achieved with  $f^{\text{Ins}}({r}^*,\bm{\pi}^*)> 0$.

\end{itemize}

Therefore, with the equal constraint of \eqref{eq:profitbound}, the premium fee $\sum_{k} \sum_{t}  \pi_k^t D_k^t$ is equal to
\begin{align}
g(r)\triangleq    -\sum_{t\in \mathcal{T}} p \cdot \mathbb{E}_{\Theta^t}[s^{t}]+c_r\cdot r
+\sum_{t\in \mathcal{T}} \mathbb{E}_{{\Theta}^t}[C_s^t(r,{\Theta^t} )]+\xi.
\end{align}

\subsection*{B. Condition $\underline{L}	<p<\overline{L}$}
We will first solve the optimal solution to \textbf{Problem-Ins} by relaxing the constraints  \eqref{eq:ICN} and \eqref{eq:IRN}, and then verify whether the solution satisfies \eqref{eq:ICN} and \eqref{eq:IRN}.

We first relax the constraints  \eqref{eq:ICN} and \eqref{eq:IRN} in \textbf{Problem-Ins}. With the equal constraint of \eqref{eq:profitbound}, we equivalently minimize
$g(r)$ to obtain the optimal solution. We can easily prove that minimizing $g(r)$ is  equivalent to minimizing the social cost in \textbf{Problem-SO}.  We  solve the optimal solution $r^*=r^\dagger$, which is the optimal capacity solution in \textbf{Problem-SO}, and the optimal $\sum_{k} \sum_{t}  \pi_k^{t*} D_k^{t*}$ is given by
$g(r^\dagger)$.

Recall that we  have relaxed the constraints  \eqref{eq:ICN} and \eqref{eq:IRN}. If the optimal $\sum_{k} \sum_{t}  \pi_k^{t*} D_k^{t*}$ satisfies \eqref{eq:ICN} and \eqref{eq:IRN}, i.e., \eqref{eq:threshold}, 
we have the optimal solution at $r^*=r^\dagger$.  Thus, we check whether $g^l(r^\dagger)\leq g(r^\dagger)\leq g^u(r^\dagger)$ is satisfied as shown in \eqref{eq:threshold}, which is equivalent to  $\underline{L}	\leq p\leq \overline{L}$.

\subsection*{C. Condition $p\leq \underline{L}$}

When $p\leq \underline{L}$, we have  $g(r^\dagger)>f^u(r^\dagger)$. The IC and IR constraints cannot be satisfied under the capacity solution $r^\dagger$ in SO. In other words, for \textbf{Problem-Ins}, the constraint \eqref{eq:IRN} at the optimum is not slack for all the hours, which must reach the lower bound or upper bound for some hours. The key to solving this problem is to discuss the binding conditions  \eqref{eq:IRN}. We prove by contradiction that when $p\leq \underline{L}$, the constraint \eqref{eq:IRN} is upper bounded at the optimum for all the hours. Note that the condition that \eqref{eq:IRN} is upper bounded for all the hours is equivalent to that \eqref{eq:IR} is upper bounded for all the hours and all the types.

First, we write the total premium fee, i.e., the objective,  as a function of $\bm{\pi}_1$ in the following 
	\begin{align}
  \sum_t\sum_ k \pi_{k}^t  D_k^t&= \sum_t  \pi_{1}^tD_a^t+  f^l(r)\\
&\leq f^u(r).
	\end{align}

Second, suppose that at the optimal solution,  $\pi_{1}^t$ is not upper bounded for all the hours. We denote the objective $\sum_t  \pi_{1}^tD_a^t+f^l(r)$ by $f^{m}(r)$ based on the given optimal binding scenarios.   We have $f^l(r)\leq f^{m}(r)<f^u(r)$. With the equality constraint \eqref{eq:profitbound}, we have $f^{m}(r^*)-g(r^*)=0$ and $ g(r^*)$ is the minimal objective value.

Third, we show the properties of some functions. All $f^l(r)$, $f^{m}(r)$,  and $f^u(r)$ are decreasing in $r$, and $g(r)$ is convex in $r$. Since $r^\dagger$ minimizes the unconstrained $g(r)$, $g(r)$ decreases over $(0,r^\dagger)$ and increases $(r^\dagger, +\infty)$. It is not difficult to show that $f^m(r)-g(r)$, $f^u(r)-g(r)$, and $f^l(r)-g(r)$   are all concave in $r$,  and they decrease over $(r^\dagger, +\infty)$.

Now, we construct the contradiction. When $p\leq \underline{L}$, we have  $f^u(r^\dagger)-g(r^\dagger)<0$, which leads to $f^m(r^\dagger)-g(r^\dagger)<0$. Since  $f^m(r^*)-g(r^*)=0$, and $f^m(r)-g(r)$ is negative and decreasing over $(r^\dagger, +\infty)$, it is straightforward that $r^*<r^\dagger$. Since the continuous functions $f^u(r^*)-g(r^*)>f^m(r^*)-g(r^*)=0$ and $f^u(r^\dagger)-g(r^\dagger)<0$, we always have a new value $r*<r'<r^\dagger$ such that $f^u(r')-g(r')=0$. Since $g(r)$ decreases over $(0,r^\dagger)$, we have  $g(r')<g(r^*)$. This contradicts that $g(r^*)$ is the minimal objective value.

Finally, with the upper-bounded constraint \eqref{eq:IRN} for all the hours, we easily have  $r^*=r^\ddagger$, which is the optimal solution to \textbf{Problem-NoIns}.

\subsection*{D. Condition $p\geq \overline{L}$}
This part proof is similar to  Subsection.C. When $p\geq \overline{L}$, we have  $g(r^\dagger)<f^l(r^\dagger)$. The IC and IR constraints cannot be satisfied under the capacity solution $r^\dagger$ in SO. In other words, for \textbf{Problem-Ins}, the constraint \eqref{eq:IRN} at the optimum is not slack for all the hours, which must reach the lower bound or upper bound for some hours. We prove by contradiction that when $p\geq \overline{L}$, the constraint \eqref{eq:IRN} is lower bounded at the optimum for all the hours. 

First, still, suppose that at the optimal solution,  $\pi_{1}^t$ is not lower bounded for all the hours. We denote the objective $\sum_t  \pi_{1}^tD_a^t+f^l(r)$ by $f^{m}(r)$ based on the given optimal binding scenarios.   We have $f^l(r)< f^{m}(r)\leq f^u(r)$. With the equality constraint \eqref{eq:profitbound}, we have $f^{m}(r^*)-g(r^*)=0$ and $ g(r^*)$ is the minimal objective value.

Third, we show the properties of some functions,  just following the same result in Subsection.C.

Now, we construct the contradiction. When $p\geq \overline{L}$, we have $f^l(r^\dagger)-g(r^\dagger)>0$, which leads to $f^m(r^\dagger)-g(r^\dagger)>0$. Since  $f^m(r^*)-g(r^*)=0$, and $f^m(r)-g(r)$ is positive at $r^\dagger$ and decreasing over $(r^\dagger, +\infty)$, it is possible that $r^*>r^\dagger$ or $r^*<r^\dagger$.
\begin{itemize}
    \item Note that if $r^*>r^\dagger$, $f^m(r)-g(r)$ decreases between $(r^\dagger,r^*)$.
        \item  If $r^*<r^\dagger$, $f^m(r)-g(r)$ increases between $(0, r^*)$.
\end{itemize}

We have the  continuous functions $f^l(r^*)-g(r^*)<f^m(r^*)-g(r^*)=0$ and $f^l(r^\dagger)-g(r^\dagger)>0$.
\begin{itemize}
    \item If  $r^*>r^\dagger$,  we always have a new value $r^\dagger<r'<r^*$ such that $f^l(r')-g(r')=0$. Since $g(r)$ increases over $(r^\dagger,r^*)$, we have  $g(r')<g(r^*)$. 
 This contradicts that $g(r^*)$ is the minimum objective value. 

    \item If $r^*<r^\dagger$, we always have a new value $r^*<r'<r^\dagger$ such that $f^l(r')-g(r')=0$. Since $g(r)$ decreases over $(r^*,r^\dagger)$, we have  $g(r')<g(r^*)$. 
 This contradicts that $g(r^*)$ is the minimum objective value. 

\end{itemize}

Finally, with the lower-bounded constraint \eqref{eq:IRN} for all the hours,  it is not difficult to show that the optimal capacity $r^*$ is equal to the optimal solution of the following concave problem.
		\begin{align*}
	\max~ &r \notag\\	
	\text{s.t.}~
	& f^{\text{No}}(r)-\sum_{t\in \mathcal{T}}{(V_1^t-p)\cdot \mathbb{E}_{\bm{\Theta}}\left[D_a^t-{s^t} \right]} \geq \xi,\\
	\text{var:}~ &r\geq0 \notag
\end{align*}   \qed

\section*{Appendix III. Proof of Proposition \ref{prop-2}}

We prove it by discussing the three electricity-bill price conditions respectively.

If  $\underline{L}	<p<\overline{L}$, the  social cost under \textbf{Ins}-utility is equivalent to social optimum, which must be no greater than that under \textbf{NoIns}-utility.   If $p\leq \underline{L}$, the  social cost under \textbf{Ins}-utility are just equal  to that under \textbf{NoIns}-utility.

If $p\geq\overline{L}$, we will first prove that the optimal capacity under SO is no greater than the under \textbf{Ins}-utility, which is further no greater than the under \textbf{NoIns}-utility. Then, we prove the social cost under under \textbf{Ins}-utility is no higher than that under \textbf{NoIns}-utility. 
\begin{itemize}
    \item (a) We prove $r^*\geq r^\dagger$. Based on the results in Appendix.II.D, considering the optimal $r^*$ under \textbf{Ins}-utility,  we have  $f^l(r^*)-g(r^*)=0$ and $f^l(r^\dagger)-g(r^\dagger)>0$. Since $f^l(r)-g(r)$ is concave,  there are actually two possible solutions, To minimize the objective with the lower-bound at \eqref{eq:IRN}, we choose the maximum one which satisfies $r^*\geq r^\dagger$. \item (b) We prove  $r^*\leq r^\ddagger$. Recall that $r^\ddagger$ is the optimal solution to \textbf{Problem-Noins}. For  $r^*$,  $f^{\text{No}}(r)-\sum_{t}{(V_1^t-p)\cdot \mathbb{E}_{\bm{\Theta}}\left[D_a^t-{s^t} \right]}$ is concave in $r$, which approaches $+\infty$ as $r\rightarrow +\infty$. Thus, $r^*$ is achieved at the maximum solution to $f^{\text{No}}(r)-\sum_{t}{(V_1^t-p)\cdot \mathbb{E}_{\bm{\Theta}}\left[D_a^t-{s^t} \right]} = \xi$. Similarly, $r^\ddagger$ is achieved at the at the maximum solution to $f^{\text{No}}(r) = \xi$. Since $f^{\text{No}}(r)-\sum_{t}{(V_1^t-p)\cdot \mathbb{E}_{\bm{\Theta}}\left[D_a^t-{s^t} \right]} \leq f^{\text{No}}(r)$, we have $r^*\leq r^\ddagger$. 
\end{itemize}
Overall, we have $r^\dagger\leq r^*\leq r^\ddagger$.
Note that social cost increases over $(r^\dagger,r^\ddagger)$ following $g(r)$ as shown in Appendix.II.C, so the social cost under under \textbf{Ins}-utility is no higher than that under \textbf{NoIns}-utility.

Based on the discussion of all the three conditions of the electricity-bill price, we complete the proof. \qed

\section*{Appendix IV. Proof of Proposition \ref{prop-3}}

To prove that  users' total cost under \textbf{InsR}-utility are no larger than that under \textbf{NoIns}-utility, we show that the \textbf{NoIns}-utility is a sub-case of \textbf{Ins}-utility depending on the electricity bill price.  Specifically, under \textbf{Ins}-utility,  if $p\leq \underline{L}$, the optimal invested capacity $r^*$ is equivalent to that under \textbf{NoIns}-utility. Furthermore, each type's IR constraint is upper bounded for each hour, which means that the cost of each user is the same as the case that he does not not purchase the insurance under the optimal invested capacity $r^*$. Thus, in terms of users' cost and invested capacity, \textbf{Ins}-utility is equivalent to \textbf{NoIns}-utility if $p\leq \underline{L}$. Since \textbf{Ins}-utility minimizes users' total premium fee (equivalent to total cost) and  $p\leq \underline{L}$ is just a sub case of all conditions,  users' total cost under \textbf{InsR}-utility is always no larger than that  under \textbf{NoIns}-utility.  \qed

\section*{Appendix V. Profit-seeking utility}

The profit-seeking utility aims to maximize her profit under the insurance contract as follows.
			\begin{align}
		\max~&f^{\text{Ins}} \\	
		\text{s.t.}~& \eqref{eq:IC} \eqref{eq:IR}, \\
			\text{var}: &r\geq0 , \bm{\pi}
	\end{align}
 
 To maximize the profit, it is straightforward that for the IR constraint  \eqref{eq:IR}, each type's premium fee at each hour will reach the upper bound. With the upper bound \eqref{eq:IR}, we can easily show that the optimal capacity under the profit-seeking \textbf{Ins}-utility is equivalent to the following problem.
 			\begin{align}
		\max~&f^{\text{No}} \\	
			\text{var}: &r\geq0,
	\end{align}
which is exactly the case that \textbf{NoIns}-utility maximizes her profit. 

Therefore, if the utility is profit-seeking, providing the insurance and not providing lead to the same renewable investment and the same profit for the utility.  They also incur the same cost for users as all the IR constraints \eqref{eq:IR} reach the upper bound  under profit-seeking \textbf{Ins}-utility.

\end{document}